\documentclass{moriond}

\def\be{\begin{equation}}
\def\ee{\end{equation}}
\def\bea{\begin{eqnarray}}
\def\eea{\end{eqnarray}}

\begin{document}
\vspace*{4cm}
\title{QUBIC EXPERIMENT}

\author{ M. STOLPOVSKIY on behalf of the QUBIC collaboration}

\address{Université Paris Diderot-Paris 7
Laboratoire APC
Bâtiment Condorcet
Case 7020 \\
75205 Paris Cedex 13, France
}

\maketitle\abstracts{
QUBIC is a ground-based experiment, currently under construction, that uses the novel bolometric interferometry technology. It is dedicated to measure the primordial B-modes of CMB. As a bolometric interferometer, QUBIC has high sensitivity and good systematics control. Dust contamination is controlled by operating with two bands -- 150 and  220 GHz. There are two possible sites for QUBIC: either Concordia station in Antarctic or in the Argentinian Puna desert. It is planned to see the first light in 2018-2019. 
}

\section{Introduction}

The QUBIC experiment is described in the white paper \cite{battistelli2011qubic}. The QUBIC abbreviation stands for "Q and U Bolometric Interferometer for Cosmology". It is a ground-based cryogenic experiment, currently in construction phase, dedicated to measure the primordial B-modes of the CMB polarisation. In the last decade scientists have been chasing the primordial B-modes. As they can only be induced by the gravitational waves produced during the inflation epoch, the B-modes are often considered as the smoking gun of inflation \cite{ade2014planck}.

QUBIC uses a novel concept called bolometric interferometry: the instrument combines the advantages of imagers and interferometers. From imagers it inherits a high sensitivity -- for the focal planes the bolometric detectors are used. These are background limited transition edge sensitive NbSi detectors with measured noise equivalent power $NEP \sim 4 \times 10^{-17} WHz^{-1/2}$. QUBIC is a millimetric equivalent of the Fizeau interferometer with 400 elements, which gives very good systematics control. QUBIC works on two frequency bands -- 150 and 220 GHz, which allows to efficiently control the dust contamination.

The QUBIC collaboration consists of more than 80 scientists from 15 institutes around the globe. The baseline site is in Antarctic at Concordia station. We also consider an alternative site in Argentina, in Puna desert. Both sites have good atmospheric conditions for CMB observations. But the seasonal changes are much stronger in Argentina and the air humidity is higher, which implies factor of 1--3 worse sensitivity on the tensor to scalar ratio $r$, depending on the observational efficiency. The difficult logistics in Antarctica would shift the timeline to about a year in comparison with Argentinian site. The approximate timeline for QUBIC is the following: in 2016 we test the focal plane and the produce the technological demonstrator, which would test the detection capabilities of the bolometric interferometry concept. Then from the end of 2016 till the end of 2017 the construction phase starts. If the Puna site will be chosen, then the installation on the site starts right away and first light will occur in 2018. In case of the Concordia station the installation and operation are postponed by one year. The final decision on the site must be made in the middle of 2016.

\begin{figure}
\begin{minipage}{0.3\linewidth}
\centerline{\includegraphics[width=1.0\linewidth]{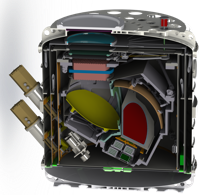}}
\end{minipage}
\hfill
\begin{minipage}{0.4\linewidth}
\centerline{\includegraphics[width=1.0\linewidth]{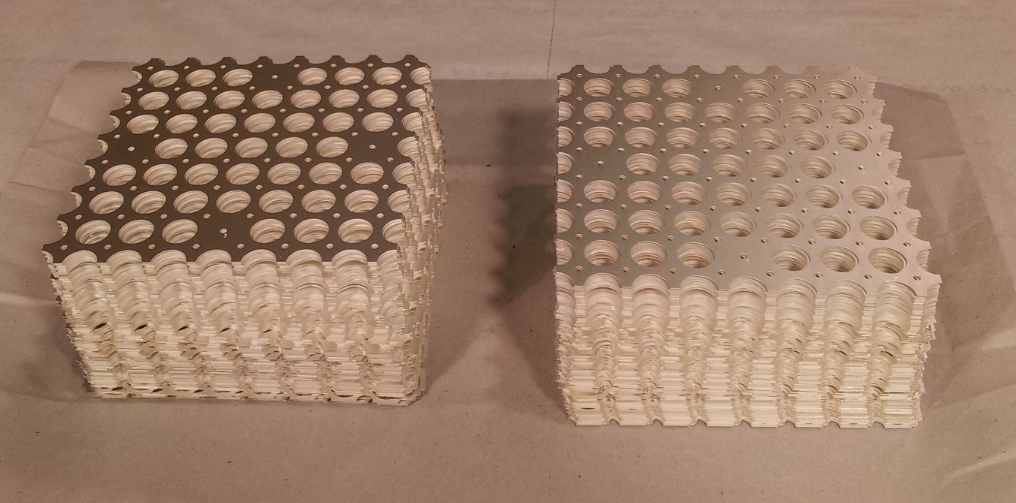}}
\end{minipage}
\hfill
\begin{minipage}{0.25\linewidth}
\centerline{\includegraphics[width=1.0\linewidth]{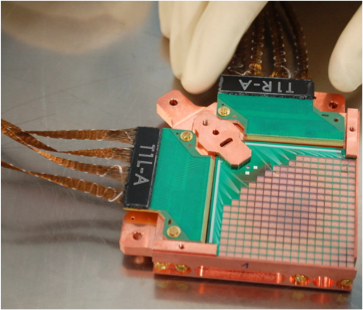}}
\end{minipage}
\caption[]{The QUBIC complete design as a 3-d model (left), the horns array for the technical demonstrator (center) and a quarter of the focal plane (right)}
\label{fig:qubic_pics}
\end{figure}

Currently the QUBIC instrument is fully designed, see figure \ref{fig:qubic_pics}, left panel. The two tubes on the left are the pulse tubes for the cryostat. Light comes through the window on the top. The pink box below the window is the horns array, which is the array of 400 pairs of horn antennas. One can consider each pair of horns as a diffractive pupil of an interferometer. The blue stripe on the horns array represents the switches that can open and close each pair of horns separately. Horns array is made of aluminium platelets. The $8 \times 8$ horns array, produced for the technological demonstrator, is shown on the middle panel of figure \ref{fig:qubic_pics}. The light from the horns is focused by two mirrors (yellow shells) on two focal planes. Mirrors act as an optical equivalents of the  correlator devices in usual interferometer concept. One of the focal planes is shown on the bottom centre of the image, another one is not shown. The light is separated by a dichroic (red panel) to two bands -- 150 and 220 GHz. One quarter of the focal plane has been produced and tested in APC, France (see right panel of figure \ref{fig:qubic_pics}). The readout system consists of time-domain SQUIDs and custom ASIC. The total multiplexing factor is 32:1 per SQUIDs and 4:1 per ASIC, thus 128:1 multiplexing in total \cite{ghribi2014latest}.

\section{QUBIC concept}

\begin{figure}[htbp]
\begin{minipage}{0.45\linewidth}
\centerline{\includegraphics[width=1.0\linewidth]{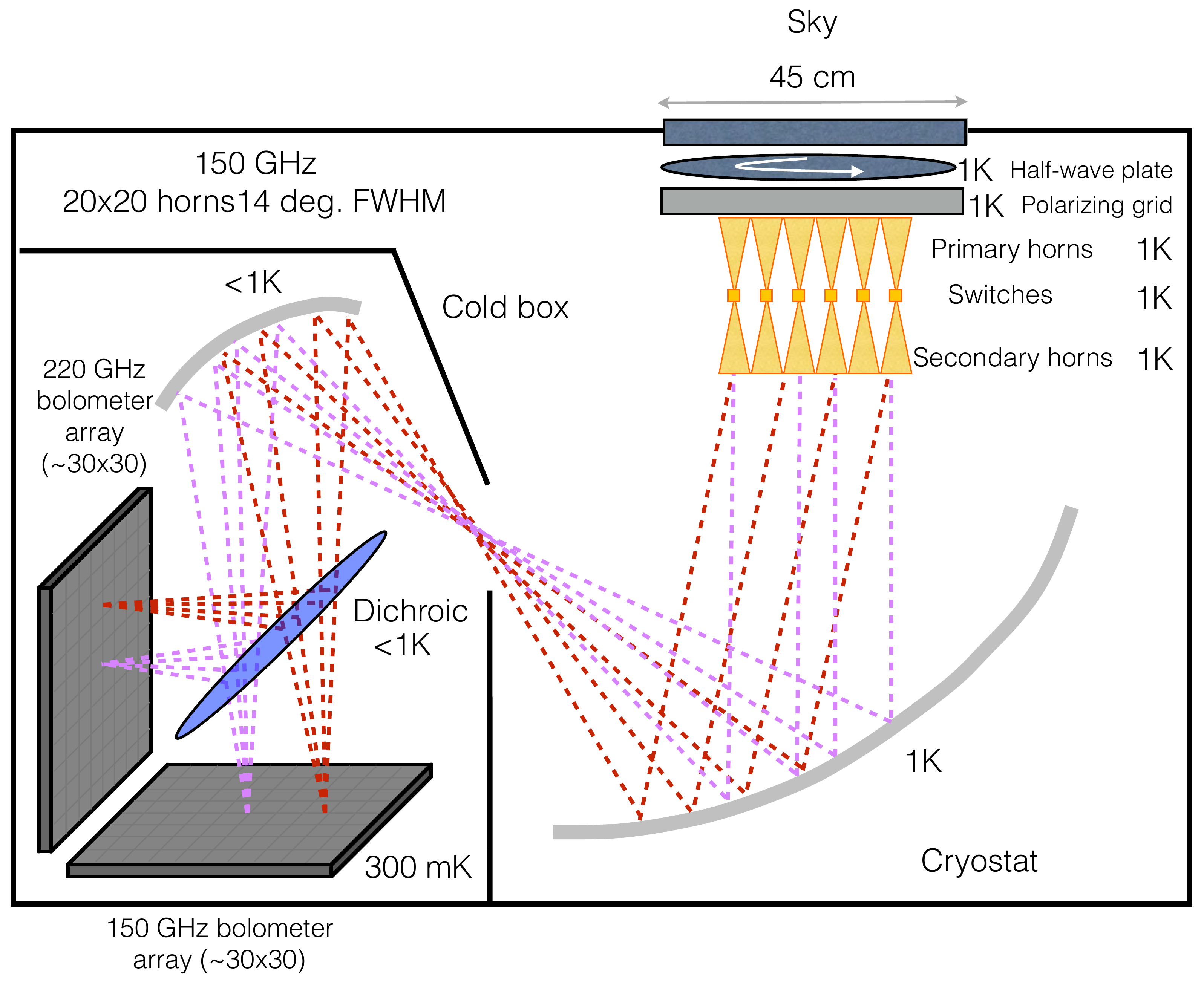}}
\end{minipage}
\hfill
\begin{minipage}{0.3\linewidth}
\centerline{\includegraphics[width=1.0\linewidth]{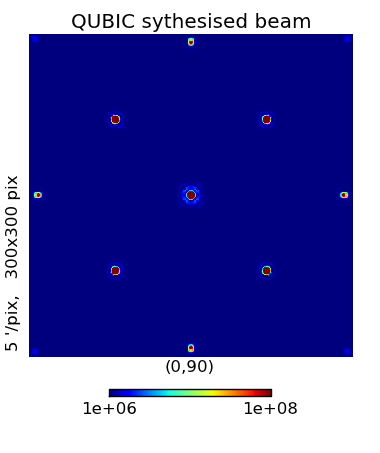}}
\end{minipage}
\caption[]{Left: schematic sketch of the QUBIC instrument. Right: QUBIC synthesised beam.}
\label{fig:sketch_sb}
\end{figure}

The QUBIC working principle is schematically shown on the picture \ref{fig:sketch_sb}, left panel. The input radiation from the sky falls on the 45 cm wide window, that also includes filters. Next to it is the cold half-wave plate, which rotates stepwise on the magnetic friction. After that there is the polarising grid and the horns array. The full width on half maximum of the primary beam is 13 degrees. The half-wave plate, polarising grid, horns array and the primary mirror are situated inside 1K cryostat. If we describe the incoming radiation in terms of electric field on two orthogonal orientations $E_x$ and $E_y$, then after the half-wave plate the radiation will be

\begin{equation}
S_{hwp} = \left(
\begin{array}{c}
    E_x cos 2 \phi(t) + E_y sin 2 \phi(t) \\
    E_x cos 2 \phi(t) - E_y sin 2 \phi(t)
\end{array} \right),
\label{eq:hwp}
\end{equation}

where $\phi(t)$ is the angle of rotation of the half-wave plate, that depends on time. The polarizing grid simply removes one component of $S_{hwp}$ and the total signal, that falls on the horns array, expressed in terms of Stocks I, Q and U parameters is $S = I + Q cos 4\phi(t) + U sin 4 \phi(t)$. This signal remains unchanged in the following components of the instrument. And even if there is cross-polarisation in, for example, horns array, it doesn't change the $S$. The QUBIC detectors are not sensitive to polarisation, thus the signal, registered by the detectors is indeed $S$. As $\phi$ depends on time, it is possible to reconstruct I, Q and U of incoming radiation. The secondary mirror and the dichroic are inside the cold box and are cooled down below 1K. The focal planes operate at temperature 300 mK.

\subsection{Self-calibration}

With 400 horn pairs the synthesised beam of QUBIC has a complex shape, shown on the figure \ref{fig:sketch_sb}. It has the main peak with full width on half maximum 23.5 arcmin and number of secondary peaks around it. The fraction of power in the secondary peaks is around 70\%. The idea of the self-calibration is the following: for a perfect instrument with zero systematics all the redundant baselines would give the same interferometric pattern on the focal planes. Using the switches one can close all the horn pairs except one baseline. One can perform measurement of an artificial point source with one baseline open. Then the same repeated with all the redundant baselines. The differences between actual interferometric patterns from redundant baselines could be fitted with systematics imperfections as parameters. As shown in the work by M-A. Bigot-Sazy et al. \cite{bigot2013self}, the systematic errors could be reduced by factor of order $10^3$ with about 30\% of time dedicated to self-calibration. 


\section{QUBIC map-making}

\begin{figure}[htb]
\begin{minipage}{0.5\linewidth}
\centerline{\includegraphics[width=\linewidth]{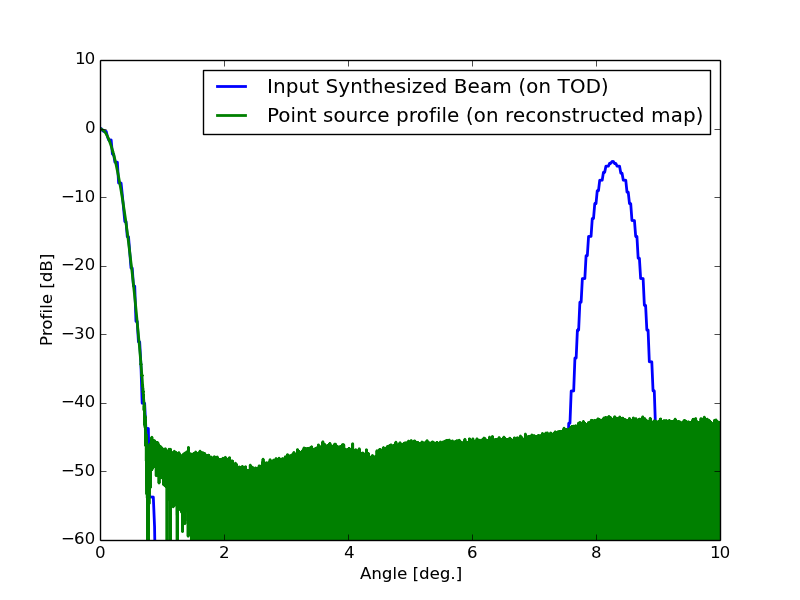}}
\end{minipage}
\hfill
\begin{minipage}{0.4\linewidth}
\centerline{\includegraphics[width=\linewidth]{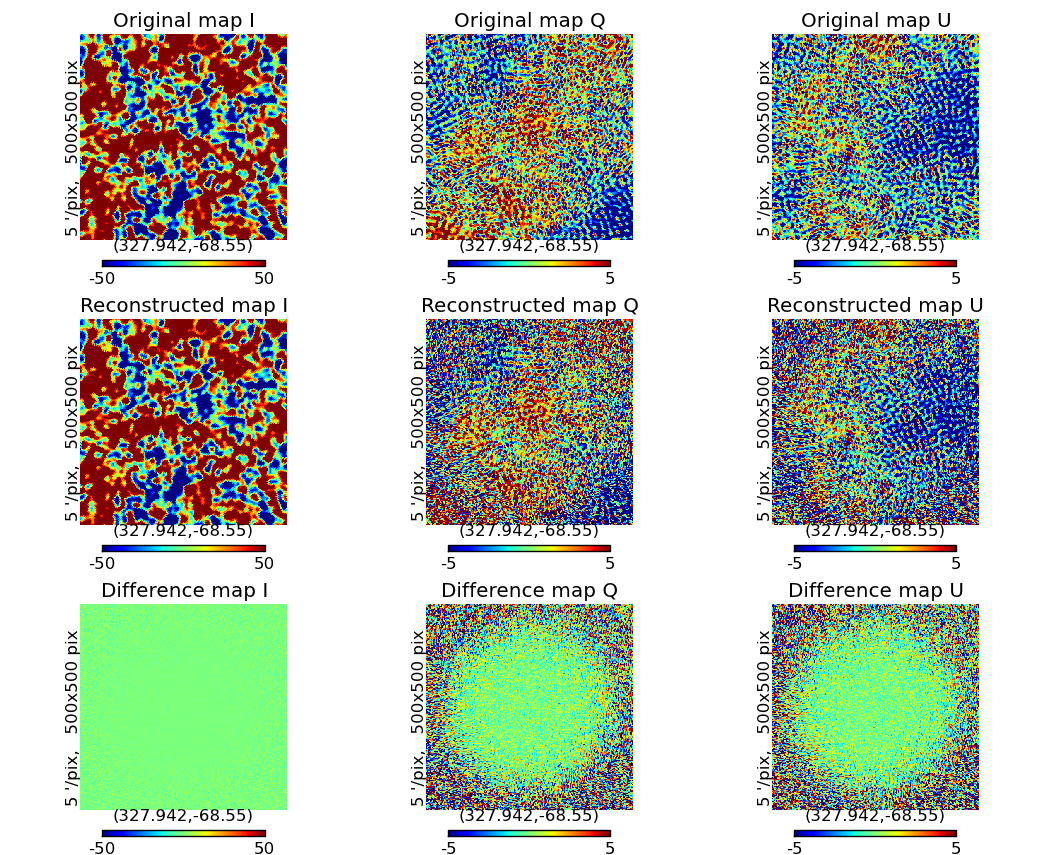}}
\end{minipage}
\caption{Left: point source observation. Blue -- direct projection of TOD to the sky, that includes all the secondary peaks. Green -- reconstruction of the original map with a single point source. Note, that the secondary peak has been deconvolved. Right: from the top row to the bottom -- the input maps, used for simulations, the reconstructed maps and the difference between the two. From the left column to the right -- the I, Q and U maps.}
\label{fig:map-making}
\end{figure}

Due to it bolometric interferometer nature, QUBIC requires a special treatment for data analysis, particularly for map-making. In the QUBIC software the synthesised beam is modelled as the sum of gaussians, centered on the positions of the synthesised beam peaks. It means, that on the time ordered data (TOD) not only the main peak is present, but also all its secondary replications. QUBIC map-making is able to resolve signal from different directions and accurately reconstructs the original signal, see fig. \ref{fig:map-making}. 

When the instrument is pointed to the edge of the coverage field, it observes signal from secondary peaks from the part of the sky, never swept by the main peak. Thus one can expect poor signal to noise ratio on the edge of the QUBIC patch. To prevent it, QUBIC map-making uses Planck maps as starting guess to reduce noise on the edge of the reconstructed map.

\section{QUBIC sensitivity}

\begin{figure}[htb]
\begin{minipage}{0.45\linewidth}
\centerline{\includegraphics[width=\linewidth]{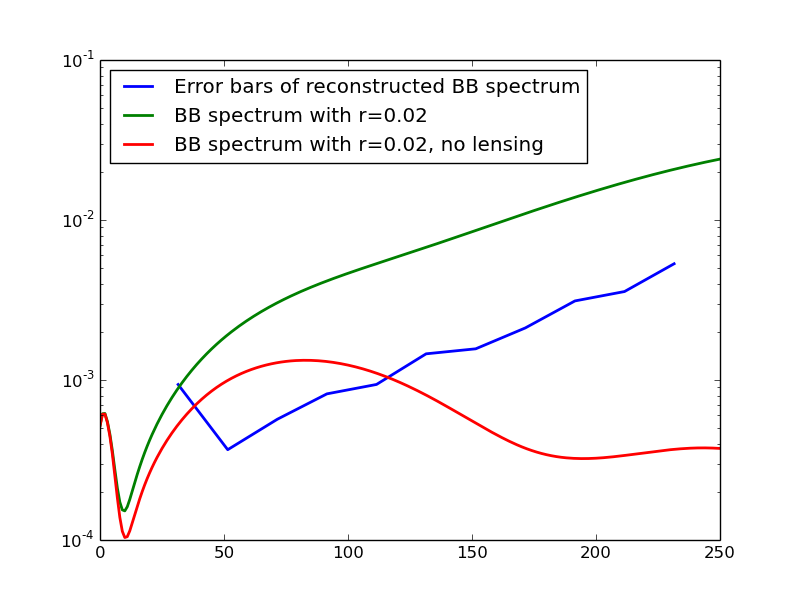}}
\end{minipage}
\hfill
\begin{minipage}{0.55\linewidth}
\centerline{\includegraphics[width=\linewidth]{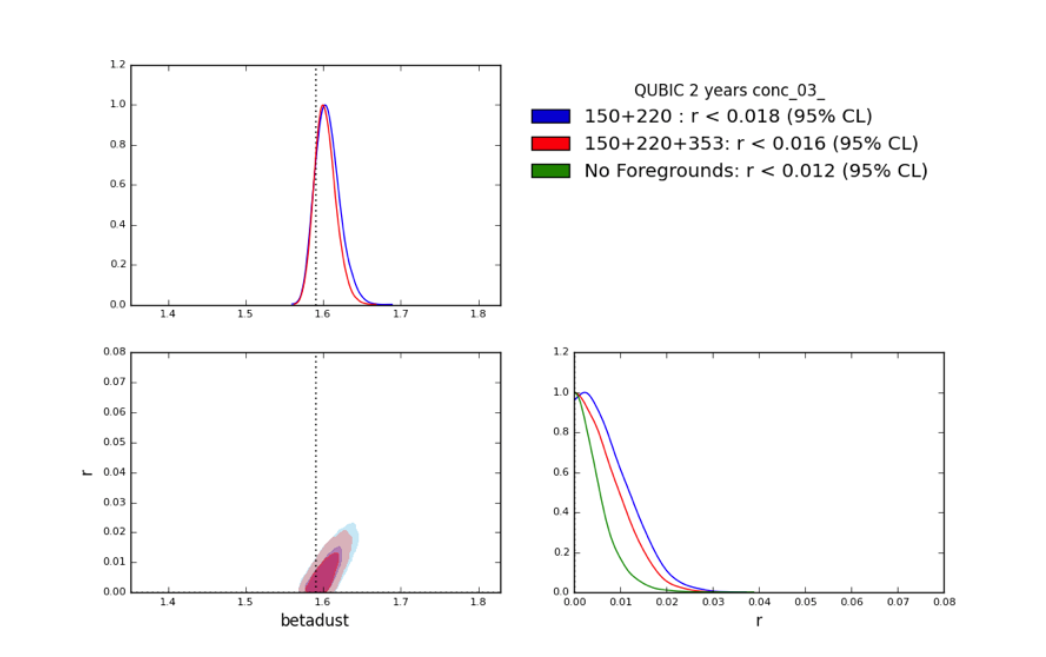}}
\end{minipage}
\caption{QUBIC sensitivity after 2 years of operation on Concordia station. Left: error-bars on BB-spectrum (blue), in comparison with BB-spectrum with $r=0.02$ with lensing (green) and without lensing (red). Right: Chi-square estimation of QUBIC sensitivity on $r$ and dust emission. The blue lines and contour is for QUBIC-alone (150 and 220 are designations for two QUBIC bands). The red lines and contour includes the Planck data of 353 GHz channel. The green line is estimated in no-foregrounds case, it is the ultimate limitation of QUBIC sensitivity.}
\label{fig:sensitivity}
\end{figure}

Running realistic simulations, that include atmospheric and detector noise, one can obtain sensitivity limits for QUBIC. It is below the peak from the primordial gravitational wave on the BB-spectrum for the value of $r = 0.02$ (see figure \ref{fig:sensitivity}, left). The Chi-square analysis for dual band observations during two years of operation, including the Planck 353 GHz data, gives constraint for sensitivity on $r < 0.016$ at 95\% confidence level assuming a single power law for dust and the level of the dust emission measured by Planck \cite{ade2015planck} (figure \ref{fig:sensitivity}, right).

\section*{References}

\end{document}